\documentclass[aip,rsi,reprint,graphicx]{revtex4-2}
\usepackage{graphicx}% Include figure files test
\usepackage{dcolumn}% Align table columns on decimal point
\usepackage{bm}% bold math
\usepackage[utf8]{inputenc}
\usepackage[T1]{fontenc}
\usepackage{mathptmx}
\usepackage{etoolbox}
\usepackage{subcaption}
\usepackage{amsmath}
\usepackage{xcolor}
\makeatletter
\def\@email#1#2{%
 \endgroup
 \patchcmd{\titleblock@produce}
  {\frontmatter@RRAPformat}
  {\frontmatter@RRAPformat{\produce@RRAP{*#1\href{mailto:#2}{#2}}}\frontmatter@RRAPformat}
  {}{}
}%
\makeatother
\begin{document}

\title{Large Errors in Kinetic Temperature Measurements Using Particle Tracking Velocimetry}
% Force line breaks with \\

\author{Anton Kananovich}
\email{kananovicha@appstate.edu}
%\homepage{http://www.Second.institution.edu/~Charlie.Author.}
\affiliation{Department of Physics and Astronomy, Appalachian State University, Boone, North Carolina 28608, USA}%
%\altaffiliation[Also at ]
\date{\today}% It is always \today, today,

\author{Parth Mehrotra}
 \affiliation{Department of Physics, Indian Institute of Science Education and Research, Dr. Homi Bhabha Road, Pune, 411008, Maharastra, India.}%Lines break automatically or can be forced with \\
  
\author{Surabhi Jaiswal}
\email{surabhi@iiserpune.ac.in}
%\homepage{http://www.Second.institution.edu/~Charlie.Author.}
\affiliation{Department of Physics, Indian Institute of Science Education and Research, Dr. Homi Bhabha Road, Pune, 411008, Maharastra, India.}%
%\altaffiliation[Also at ]
\date{\today}% It is always \today, today,

\begin{abstract}
	We report on random errors in kinetic temperature measurements due to finite spatial resolution in particle tracking velocimetry. Using simulated data, we isolate the error caused by finite spatial resolution from other sources of uncertainty, such as particle acceleration and particle mismatch. A sample of particle velocities is generated from a Maxwellian distribution at a prescribed kinetic temperature. Particle positions are assigned randomly and discretized to match a prescribed spatial resolution. Velocities are reconstructed using the two-frame tracking method, and the resulting kinetic temperature is calculated and compared to the true kinetic temperature. Results show that under typical experimental conditions, the uncertainty in particle positions propagates into large errors in the velocity distribution and the measured kinetic temperature. We find that this might introduce errors ranging from tens of percent at high kinetic temperatures ($\sim 10$~eV) to thousands of percent at low temperatures ($\sim 0.1$~eV).
\end{abstract}

\maketitle

\section{\label{sec:level1}Introduction}

As is well known in statistical physics, macroscopic properties of matter—volume, pressure, temperature, flow speed, and others—are determined by the microscopic behavior of the particles that compose it. Ideally, one would like to measure these microscopic properties directly, but in practice this is rarely possible. 

Direct access to microscopic particle properties is possible only in special cases. In simulations, molecular dynamics (MD) simulations provide complete access to individual particles properties. In experiments, systems such as granular materials and dusty plasmas allow individual particles to be resolved. In these situations, video microscopy can be used to measure particle positions frame by frame, from which trajectories, velocities, and even accelerations can be reconstructed. This approach of extracting particle velocities from video data is known as Particle Tracking Velocimetry (PTV)~\cite{crocker1996methods,feng2007accurate,feng2011errors,mohr2019algorithms}.

Although the microscopic information obtained by PTV can, in principle, be used to determine macroscopic quantities of interest, one must account carefully for the uncertainties introduced at each stage. Uncertainty in particle positions, as determined from PTV, propagates directly into uncertainty in particle velocities. In turn, uncertainty in particle velocities propagates into the macroscopic thermodynamic quantities derived from them.

Three main sources of these uncertainties have been previously identified. The first arises from the finite spatial resolution of the imaging system, which limits the precision of particle position measurements~\cite{feng2011errors}. The second arises from particle acceleration during the sampling interval; the standard two-frame tracking method assumes straight-line motion at constant velocity, which leads to errors when the velocity changes~\cite{feng2011errors}. The third arises from tracking mismatch, where particles are incorrectly identified between frames~\cite{kantor2014bias}.

In this paper, we analyze the first type of error, caused by finite spatial resolution. Previous attempts to estimate this error in kinetic temperature measurements and velocity autocorrelation functions encountered two main difficulties~\cite{feng2011errors}. The true kinetic temperature of a real experimental system is usually unknown, and the contributions of the two error sources are intertwined. In Ref.~\cite{feng2011errors}, the authors could not quantify this discrepancy. They concluded that PTV does not yield a unique value for the kinetic temperature, but instead a value that depends on the sampling interval.

We present a method to quantify the error in kinetic temperature measurements using PTV due to finite camera resolution. This method isolates the error arising from finite spatial resolution from the error due to particle acceleration. We prescribe a velocity distribution function for a known kinetic temperature with arbitrary precision, and we compare this true value to the result produced by the PTV algorithm. This approach eliminates the two difficulties identified previously~\cite{feng2011errors}: the true kinetic temperature is known by design, and the error due to finite resolution is isolated. Our method is not based on Molecular Dynamics (MD) simulations; this avoids artifacts specific to MD.

Using this approach, we obtain lower-bound estimates for the errors in kinetic temperature measurements by PTV. We evaluate these uncertainties for parameters typical of dusty plasma experiments and find that the errors are significantly larger than previously anticipated. We also outline strategies to reduce these errors. We also confirm the counterintuitive observation reported by Feng et al.~\cite{feng2011errors} that the use of higher-frame-rate cameras can, under certain conditions, increase rather than decrease the error in kinetic temperature.

These findings apply generally to PTV measurements in any physical system, including granular materials~\cite{jain2002experimental,tsai2004slowly}, colloids~\cite{crocker1996methods,leocmach2013novel}, fluid dynamics~\cite{luethi2005lagrangian,dore2009investigation}, and biophysics~\cite{lima2009measurement}. In this work, we specifically evaluate these errors for conditions typical of dusty plasma experiments~\cite{piel2010dusty,fortov2007physics,shukla2001introduction,tsytovich2008elementary,vladimirov2005physics,konopka2016guest,goree2008fundamentals,melzer2019physics}.

\section{\label{sec:level2}Methodology}

\begin{figure}[ht]
	\includegraphics[width=0.85\columnwidth]{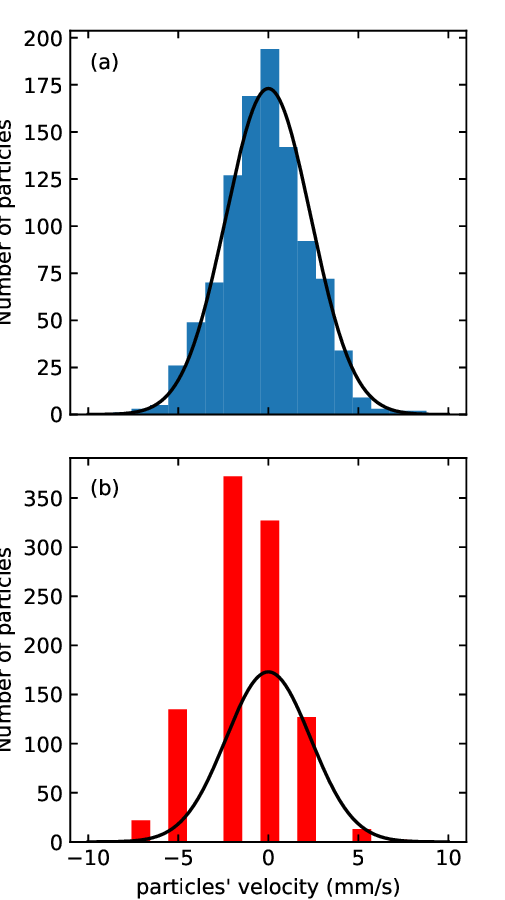}
	\caption{\label{fvelHist}
		Comparison of true and measured velocity distributions. (a) Histogram of a random velocity sample drawn from Eq.~(\ref{eqMax}) (Methodology Step 1). (b) Histogram of velocities reconstructed from the same sample using the PTV method (Methodology Step 5) for resolution $R = 24.39\,\mu\mathrm{m/px}$ and a frame rate of $99$~fps. The solid line in both panels represents the theoretical Maxwellian distribution. Note that in panel (b), the velocities take on only eight discrete values. This is a phenomenon peculiar to PTV called "pixel locking"~\cite{feng2007accurate}. The discrepancy between the Maxwellian distribution and the shape of the histogram in (b) illustrates the error introduced by finite spatial resolution. Results are shown for a true kinetic temperature of 10~eV.}
\end{figure}

Our goal is to determine how spatial resolution and sampling interval alone produce errors in kinetic temperature when using the standard two-frame PTV procedure. To isolate these contributions, we construct a simulation in which the true kinetic temperature is prescribed exactly, and all deviations arise solely from the PTV analysis step.

We do not use MD simulations for this purpose. MD introduces several complications that are undesirable here. First, a thermostat maintains the temperature only within fluctuations that depend on MD parameters, adding an uncertainty unrelated to PTV. Second, MD includes interparticle collisions and the resulting accelerations, which would reintroduce acceleration-related errors that we specifically wish to exclude. Third, the particles systems like dusty plasmas are known to deviate from Gaussian statistics~\cite{ratynskaia2005statistics,liu2008non,ott2009is,andrew2025anisotropic,melzer2025metal}, which complicates the kinetic-theory definition of temperature. Our method avoids these issues by prescribing the velocity distribution directly.

The simulation is based on simple first principles. We prescribe both the kinetic temperature and the corresponding velocity distribution. For a chosen one-dimensional temperature $T_x$, the Maxwellian distribution is a Gaussian with variance $k_B T_x/m$. We then generate a random sample of particle velocities drawn from this prescribed distribution. In this way, the true kinetic temperature is known precisely, and any discrepancy found in the PTV-derived temperature arises solely from the PTV analysis due to the imposed spatial resolution and sampling interval.

\begin{figure}[t]
	\includegraphics[width=0.95\columnwidth]{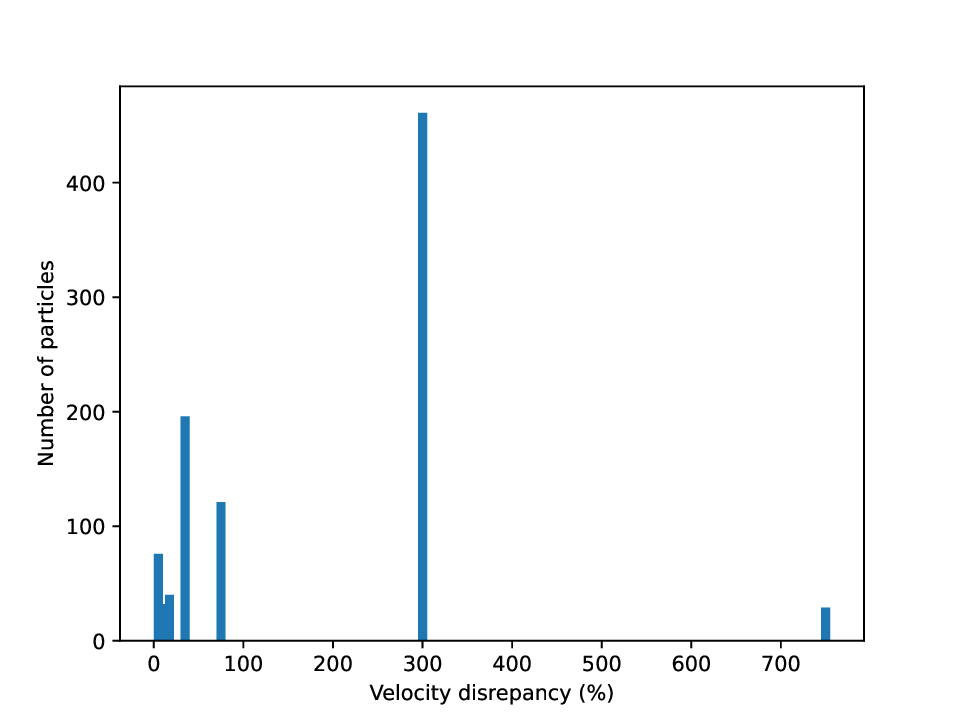}
	\caption{\label{fHist}
		Histogram of discrepancies between true particle velocities $v_{xi}$ and velocities reconstructed using PTV $v_{x i,\mathrm{meas}}$. The figure illustrates that roughly half of the reconstructed velocities deviate from the true values by 300\% or more. These large microscopic errors inevitably propagate into the macroscopic calculation of the kinetic temperature. The data correspond to the same random sample used in Fig.~\ref{fvelHist}, drawn from a Maxwellian distribution at a kinetic temperature of 10~eV. Discrete values of the discrepancies are the manifestation of the "pixel locking"~\cite{feng2007accurate}.}
\end{figure}

The simulation and PTV analysis proceed as follows.

\begin{figure*}[ht]
	\includegraphics[width=1\textwidth]{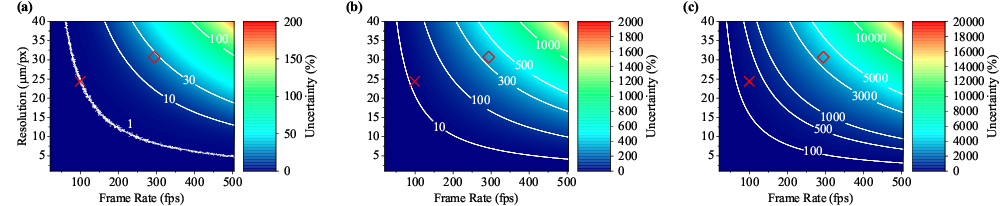}
	\caption{\label{fContour}
Contour plots of the fractional error in the kinetic temperature measured by PTV, as a function of camera frame rate and spatial resolution. Results are shown for three true kinetic temperatures: (a) 10~eV, (b) 1.0~eV, and (c) 0.10~eV. The error arises solely from finite spatial resolution. The red cross (\textcolor{red}{$\times$}) marks the operating point for our first experimental condition ($\nu=99~\mathrm{fps}$, $R=24.39~\mu\mathrm{m/px}$), and the red diamond (\textcolor{red}{$\diamond$}) marks the second condition ($\nu=294~\mathrm{fps}$, $R=30.69~\mu\mathrm{m/px}$). The white contour lines represent isolines of constant fractional error. The uncertainty increases with frame rate and decreases with coarser spatial resolution (larger $R$). Note the large uncertainties for typical frame rates and resolutions (center of each panel), especially in the low-temperature case (panel (c)). }
\end{figure*}

\begin{enumerate}
	
	\item \textit{Prescribing the temperature and velocities.}  
	For a prescribed one-dimensional kinetic temperature $T_x$, the Maxwellian velocity distribution is
	\begin{equation} \label{eqMax}
		f(v_x) = \sqrt{ \frac{m}{2\pi k_B T_x}} \exp{\!\left(-\frac{m v_x^2}{2 k_B T_x}\right)},
	\end{equation}
	where $m$ is the particle mass and $k_B$ the Boltzmann constant.  
	Using a random number generator, we create a sample $v_{x1}, v_{x2}, \ldots, v_{xN}$ drawn from Eq.~(\ref{eqMax}). This models an experimental particle cloud of size $N$, where the $i$th particle has velocity $v_{xi}$.
	
	\item \textit{Assigning particle positions.}  
	Initial particle positions $x_{\mathrm{prev},i}$ are drawn from a uniform distribution between $x_{\min}$ and $x_{\max}$. Thus, each particle has a velocity $v_{xi}$ and an initial position $x_{\mathrm{prev},i}$.
	
	\item \textit{Propagating to the next frame.}  
	For a chosen frame rate $\nu$, the sampling interval is $\Delta t = 1/\nu$.  
	The true continuous position of particle $i$ in the next frame is
	\begin{equation}
		x_{\mathrm{next},i} = x_{\mathrm{prev},i} + v_{xi}\,\Delta t.
	\end{equation}
	No interparticle forces or accelerations are included.
	
	\item \textit{Rounding positions to a pixel grid.}  
	The camera resolution is characterized by a scale factor $\alpha$ in units of pixels per meter. To simulate finite spatial resolution, the true positions are scaled by $\alpha$, rounded to the nearest integer, and scaled back to SI units:
	\begin{equation}
		x_{\mathrm{prev},i}^{(\mathrm{px})} = \frac{\mathrm{round}(x_{\mathrm{prev},i}\,\alpha)}{\alpha},
	\end{equation}
	\begin{equation}
		x_{\mathrm{next},i}^{(\mathrm{px})} = \frac{\mathrm{round}(x_{\mathrm{next},i}\,\alpha)}{\alpha}.
	\end{equation}
	This discretization introduces an effective position uncertainty of one pixel. Both frames are sampled on the same pixel grid, and the only loss of information comes from this rounding.
	
	\item \textit{Reconstructing velocities using the two-frame PTV algorithm.}  
	The measured velocity of particle $i$ is calculated using the standard two-frame PTV formula~\cite{feng2011errors,feng2016particle}:
	\begin{equation}\label{eqV}
		v_{x i,\mathrm{meas}} = \frac{x_{\mathrm{next},i}^{(\mathrm{px})} - x_{\mathrm{prev},i}^{(\mathrm{px})}}{\Delta t}.
	\end{equation}
	Thus, the only sources of error in $v_{x i,\mathrm{meas}}$ are the finite pixel size and the chosen sampling interval.
	
	\item \textit{Computing the measured kinetic temperature.}  
	From the measured velocities we compute the kinetic temperature
	\begin{equation}
		T_{\mathrm{meas}} = \frac{m}{k_B}\,\langle v_x^2 \rangle,
	\end{equation}
	where the variance is
	\begin{equation}
		\langle v_x^2 \rangle = \frac{1}{N}\sum_{i=1}^{N} 
		\left( v_{x i,\mathrm{meas}} - \overline{v}_{x,\mathrm{meas}} \right)^2 .
	\end{equation}
	
	\item \textit{Estimating the error in kinetic temperature.}  
	Steps~1–6 are repeated $N_{\mathrm{rep}} = 100$ times. The uncertainty in the measured temperature is estimated as the standard deviation of the mean,
	\begin{equation}
		\delta T = \frac{\sigma_T}{\sqrt{N_{\mathrm{rep}}}},
	\end{equation}
	where
	\begin{equation}
		\sigma_T = \sqrt{\frac{1}{N_{\mathrm{rep}}} 
			\sum_{k=1}^{N_{\mathrm{rep}}} \left( T_x - T_{\mathrm{meas},k} \right)^2 } .
	\end{equation}
	
\end{enumerate}

\section{Discussion}

We applied the method described above to estimate the error in the kinetic temperature measured by PTV as a function of camera frame rate and spatial resolution. All numerical values were chosen to match conditions typical of laboratory dusty plasma experiments, so that the results can be compared directly to experimental practice.

For these calculations we used a sample size of $N = 1000$, which is a common number of microspheres in a two-dimensional particle cloud~\cite{smith2008phase,haralson2016laser,haralson2017overestimation,couedel2018full,kananovich2020experimental,kananovich2020shocks,mendezharper2020origin,vasilieva2021laser,kananovich2021shock,alekseevskaya2023isotropic,huang2023dissipative}. The particle mass was $m = 2.9\times10^{-13}\,\mathrm{kg}$, corresponding to melamine--formaldehyde microspheres of diameter $7.14\,\mu\mathrm{m}$, which are also used in our laboratory.

Finite spatial resolution distorts the measured velocity distribution, as shown in Figures~\ref{fvelHist} and \ref{fHist}. In panel (a) of Fig.~\ref{fvelHist}, we show the histogram for a random sample of $N=1000$ velocities drawn from Eq.~(\ref{eqMax}) (Step 1 of the Methodology) corresponding to a kinetic temperature of $T_x = 10$~eV. The theoretical Maxwellian distribution is shown as a solid line. In panel (b), we show the histogram for the \textit{same sample}, but obtained from velocities reconstructed using the PTV method (Step 5 of the Methodology) for a camera with resolution $R = 24.39\,\mu\mathrm{m/px}$ and a frame rate of $99$~fps. Note that this histogram has only eight non-empty bins. This is a well-known phenomenon called "pixel locking"~\cite{feng2007accurate}. Since we can obtain coordinates from the pixel grid only as discrete values, dividing their difference by the time interval in Eq.~(\ref{eqV}) yields discrete velocities. These errors in individual particle velocities propagate directly into the kinetic temperature; this is evident in panel (b), where the shape of the reconstructed histogram deviates significantly from the solid line representing the true Maxwellian distribution.

Figure~\ref{fHist} further illustrates the error in particle velocities. Here, we aggregated the discrepancies between the true and measured velocities for the same data used in Fig.~\ref{fvelHist}. Most of the particles (for these camera parameters) have very large errors. The values of these errors are also discrete; this is a manifestation of the same pixel-locking phenomenon responsible for the empty bins in Fig.~\ref{fvelHist}(b).

Our main results are shown in Fig.~\ref{fContour}. Fig.~\ref{fContour} presents three contour plots showing the discrepancy in the measured kinetic temperature for true temperatures of 10~eV, 1.0~eV, and 0.10~eV. These values span the range commonly encountered in dusty plasma experiments. To illustrate how the results apply to an actual experiment, we mark two points corresponding to our laboratory conditions. The red cross (\textcolor{red}{$\times$}) indicates operation at $R = 24.39\,\mu\mathrm{m/px}$ and frame rate $99$~fps, while the red diamond (\textcolor{red}{$\diamond$}) indicates operation at $R = 30.69\,\mu\mathrm{m/px}$ and frame rate $294$~fps.

Two general trends are visible in all three panels. First, for a fixed frame rate, the error decreases for cameras with coarser resolution (larger values of $R$). This follows directly from the fact that discretization of particle positions is the only source of error included in our model. For example, in panel (a), the uncertainty corresponding to our point at $294$~fps and $30.69\,\mu\mathrm{m/px}$ is about $26\%$, whereas for the same frame rate but the coarser resolution of $5\,\mu\mathrm{m/px}$ the uncertainty is below $10\%$.

Second, for a fixed spatial resolution, the uncertainty increases with frame rate. In panel (a), the error for our point at $99$~fps and $24.39\,\mu\mathrm{m/px}$ is about $1\%$, but if the frame rate were increased to $400$~fps, the error would exceed $30\%$. Although this trend may appear counterintuitive, it was previously reported by Feng \textit{et al.}~\cite{feng2011errors}. The reason is that at high frame rates the displacement of a microsphere between frames becomes smaller than one pixel, causing the measured velocity to approach zero regardless of the true velocity. This error in particle velocities then propagates directly into the kinetic temperature.

The main result of this paper is the magnitude of these uncertainties. For typical experimental parameters (corresponding to the center of each panel), the errors are tens of percent at high temperatures (panel (a)), hundreds of percent at intermediate temperatures (panel (b)), and thousands of percent at low temperatures (panel (c)). Such large errors imply that PTV yields unreliable kinetic temperatures for low-temperature conditions unless the spatial resolution and sampling interval are chosen with great care. We emphasize that our calculations include only the contribution from finite spatial resolution. Errors arising from particle acceleration were not included. Therefore, the uncertainties shown in Fig.~\ref{fContour} represent a lower bound.

We outline a strategy to estimate and reduce these errors. First, the experimenter should quantify the uncertainty for their specific setup using the methodology in Sec.~\ref{sec:level2}. Simulating the PTV process with specific hardware parameters yields the baseline error due to spatial resolution. Second, if this error is too large, it can be reduced by skipping frames during analysis to increase the effective sampling interval. The simulation methodology can explicitly predict the improvement in accuracy gained by this adjustment. We caution, however, that this estimate represents a lower bound. It isolates the effect of finite resolution. While skipping frames reduces resolution-based errors, it may exacerbate errors arising from particle acceleration and tracking mismatches.

\section{Summary}

In summary, we have isolated and quantified the random errors in kinetic temperature measurements arising from finite spatial resolution in PTV. We demonstrated that "pixel locking"—a phenomenon where the measured velocity distribution collapses into discrete values—causes large random errors in individual particle velocities. These errors propagate into the uncertainty in the macroscopic kinetic temperature.

We found that the magnitude of this uncertainty is significant. Under typical conditions for dusty plasma experiments, the measured temperature can deviate from the true value by hundreds or thousands of percent. We confirmed previous findings~\cite{feng2011errors} that increasing the camera frame rate—often done to improve temporal resolution—can actually increase this error. This occurs because at high frame rates, the particle displacement between frames becomes comparable to the pixel size, making the discretization error dominant.

Finally, we provided a strategy to diagnose and mitigate these errors. Experimenters can use the simulation methodology presented here to calculate the lower-bound uncertainty for their specific hardware parameters. If this baseline error is excessive, it can be reduced by analyzing the data with a larger effective sampling interval (skipping frames), thereby ensuring that particle displacements are sufficiently resolved compared to the pixel size.

\section{Acknowledgements}
Appalachian State University was supported by the United States Department of Energy under Grant No. DE-SC0025444 and the National Science Foundation under Grant No. PHY-2510502. Work at  IISER Pune was supported by IISER internal startup grant. 

\section{References}

%\bibliography{references.bib}% Produces the bibliography via BibTeX.

%aipnum4-2.bst 2019-01-14 (MD) hand-edited version of apsrev4-1.bst
\providecommand{\noopsort}[1]{}\providecommand{\singleletter}[1]{#1}%

\end{document}